\documentstyle[12pt,blois,psfig]{article}

\begin{document}
\newcommand{\hI}{\mbox{${\rm H\ I}$}}
\newcommand{\lya}{\mbox{${\rm Ly}\alpha$}}
\def\cge      {{$_ >\atop{^\sim}$}}
\def\cle      {{$_ <\atop{^\sim}$}}
\def\hst      {{\sl HST}}
\def\wfpc     {{\sl WFPC2}}
\def\etal     {{\it et\thinspace al.} }
\def\spt      {{$\buildrel{s}           \over .$}}
\def\eg       {{\it e.g.}}
\def\err#1#2 {{$_{#1}\atop{^{#2}}$}}

\heading{THE ULTRAVIOLET LUMINOSITY DENSITY OF THE UNIVERSE FROM PHOTOMETRIC
REDSHIFTS OF GALAXIES IN THE HUBBLE DEEP FIELD}

\author{S. M. Pascarelle \& K. M. Lanzetta $^{1}$, A. Fern\'andez-Soto $^{2}$}
{$^{1}$ SUNY Stony Brook, Stony Brook, USA.}  {$^{2}$ University of New South
Wales, Sydney, Australia.}

\begin{bloisabstract}
Studies of the Hubble Deep Field (HDF) and other deep surveys have revealed an
apparent peak in the ultraviolet (UV) luminosity density, and therefore the
star-formation rate density, of the Universe at redshifts $1<z<2$. We use
photometric redshifts of galaxies in the HDF to determine the comoving UV
luminosity density and find that, when errors (in particular, sampling
error) are properly accounted for, a flat distribution is statistically
indistinguishable from a distribution peaked at $z\simeq 1.5$.
Furthermore, we examine the effects of cosmological surface brightness (SB)
dimming on these measurements by applying a uniform SB cut to all galaxy
fluxes after correcting them to redshift $z=5$. We find that, comparing all
galaxies {\em at the same intrinsic surface brightness sensitivity}, the UV
luminosity density contributed by high intrinsic SB regions increases by
almost two orders of magnitude from $z\simeq 0$ to $z\simeq 5$.
This suggests that there exists a population of objects
with very high star formation rates at high redshifts that apparently do
not exist at low redshifts. The peak of star formation, then, may occur
somewhere beyond a redshift $z$\cge 5.
\end{bloisabstract}

\section{Introduction}


The ultraviolet (UV) luminosity density was measured at redshifts $z$\cle 1
by Lilly \etal (1996) \cite{L96} using the Canada--France Redshift
Survey (CFRS) and found to rise rapidly by a factor of $\sim 15$ from
$z=0$ to $z=1$. Similar results were found in this redshift range by
\cite{CO97}. Madau, Pozzetti, \& Dickinson (1998) \cite{M98} measured the UV
luminosity density at redshifts $z\simeq 2.75$ and $z\simeq 4.00$ using the
$U$- and $B$-band ``dropout'' technique applied to galaxies in the Hubble Deep
Field (HDF) and found that it might be lower than the
\cite{L96} measurements at $z$\cle 1 by a factor of $\sim 2$.
This suggested that there might be a peak in the UV luminosity density
at some redshift between $z = 1$ and 2.

More recently, Connolly \etal (1997) \cite{C97} measured the
UV luminosity density from the HDF in the previously unsurveyed redshift range
1\cle $z$ \cle 2 and found that it appears to peak at $z \simeq 1.5$, which is
consistent with the results of \cite{L96} and \cite{M98}. However the
photometric redshifts of galaxies in the HDF from Sawicki, Lin, and Yee (1997)
\cite{SLY97} result in a UV luminosity density
that continues to increase to $z$\cge 2.5, indicating that the peak at $z
\simeq 1.5$ may be questionable.

\section{Photometric Redshifts}

  The starting point of our analysis is the photometric redshift estimates of
\cite{LYF96} and \cite{FLY98}. The details of the photometric redshift
estimation technique have been and will be presented elsewhere (\cite{LFY98} \&
\cite{FLY98}). To briefly summarize, galaxy photometry is determined from the
optical F300W, F450W, F606W, and F814W \cite{W96} and infrared $J$, $H$, and
$K$ \cite{D98} images of the HDF.
Galaxy redshifts are then determined from fits to the spectral templates of
E/S0, Sbc, Scd, and Irr galaxies,
including the effects of intrinsic and intervening
neutral hydrogen absorption.
The result of the most recent application of this method is presented in the
catalog of \cite{FLY98}, which lists photometric redshift estimates
of 1067 galaxies to a limiting magnitude threshold of $AB(814) = 28.0$ over an
angular area of $\sim$4 arcmin$^2$.
Comparison between the spectroscopic and photometric redshifts, as shown in
Figure 1, indicates that the agreement is excellent at $z <  2$ ($\sigma =
0.09$) and remains quite good all the way out to $z$\cge 5 ($\sigma\leq 0.45$).

\begin{figure}[h]
\centerline{\psfig{file=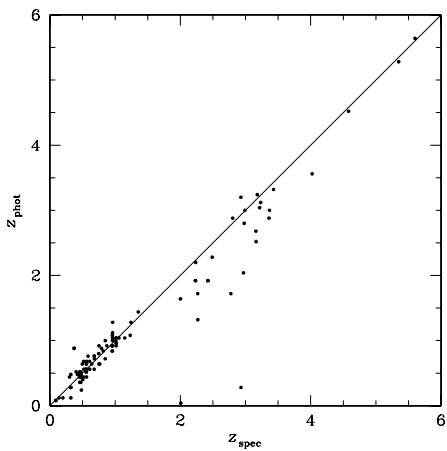,width=6.2cm}}
\caption{\small{Comparison of photometric and spectroscopic redshifts of
galaxies in the HDF}}
\end{figure}


\section{Ultraviolet Luminosity Density}

  We determined the luminosity (per unit wavelength interval) of each
galaxy at a rest-frame wavelength $\sim$1500\AA\ by applying an empirical
$K$-correction derived from the best-fit spectral template to the measured
galaxy photometry.
From this, we determined the comoving luminosity density versus redshift.
The key point of our measurement is that we
determined the uncertainty of the comoving luminosity density versus redshift
by applying a bootstrap resampling technique.  For each iteration of the
bootstrap technique, we resampled the photometric catalog and redetermined the
photometric redshift of each resampled galaxy, perturbing the photometry by
random deviates according to the measured photometric error and perturbing the
redshift by a random deviate according to the RMS residuals described in \S 2.
We then determined the comoving luminosity density versus redshift using the
resampled, perturbed redshift catalog.  We repeated the procedure 1000 times
in order to determine the range of values compatible with the observations.
This procedure explicitly allows for sampling error, photometric error, and
cosmic variance with respect to the spectral templates.

\begin{figure}[t]
\centerline{\psfig{file=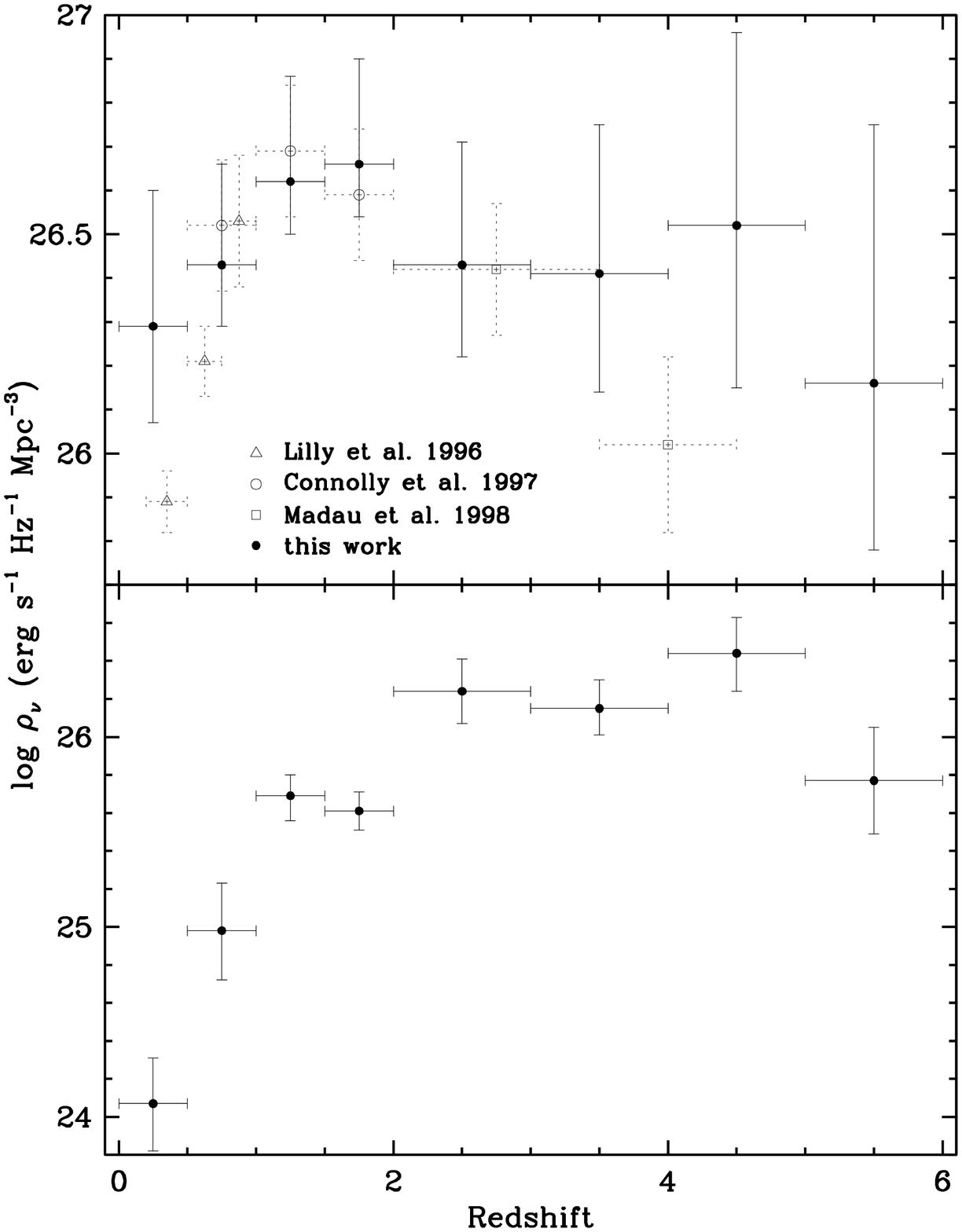,width=9.0cm}\ \ \
\psfig{file=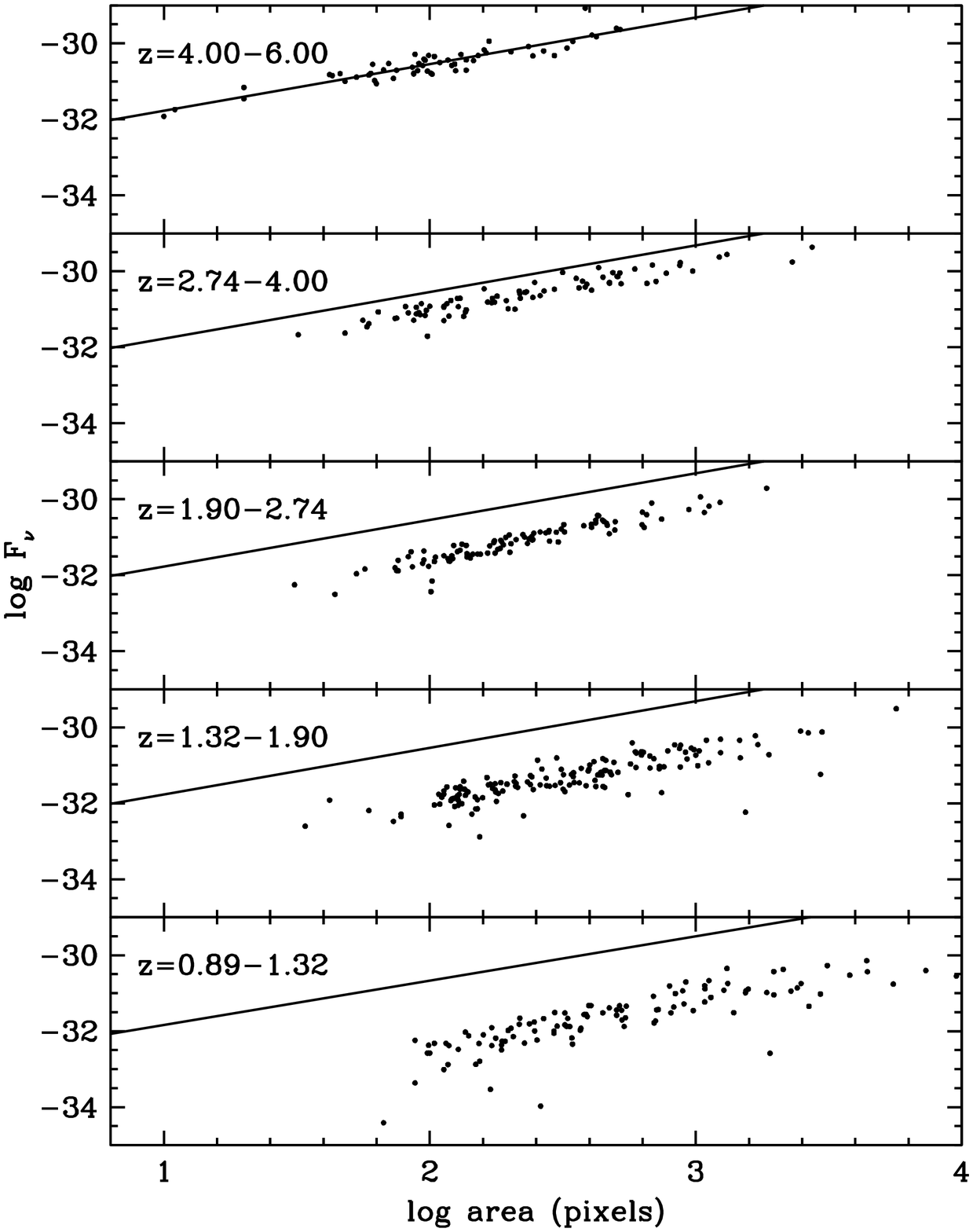,width=9.0cm}}
\caption{\small{({\it left panel}), The UV luminosity density as a function of
redshift as measured from galaxies in the HDF using our photometric redshifts
({\it top}).
Data points from \cite{L96} (open triangles), \cite{C97} (open circles), and
\cite{M98} (open squares) are included
for comparison with our measurements (filled circles). 
The UV luminosity density arising from intrinsically high surface brightness
(SB) regions, after applying a uniform SB cut is shown in the bottom panel.
({\it right panel}), Flux versus area (which is essentially average surface
brightness) for galaxies in the HDF. Note that it appears that
galaxies have lower overall surface brightnesses at redshifts below $z=4$ than
at $z=4-6$.}}
\end{figure}

The results are shown in the top left panel of Figure 2, which plots the
comoving UV luminosity density versus redshift. For comparison, the data
points from \cite{L96}, \cite{C97}, and \cite{M98} are also plotted. Note that
our data are entirely consistent with the
previous measurements to well within $\simeq 1.5\sigma$ (and all but our first
data point are within \cle1$\sigma$ of previous data).

With sampling errors properly accounted for, the errorbars generated from our
bootstrap code are more than a factor of two larger than those of \cite{C97}
or \cite{M98} for the $z>2$ UV luminosity density. 
It can be seen that, after an increase at redshifts $z \simeq 0 - 2$,
the UV luminosity density remains constant to within errors to redshifts
$z\simeq 6$. In other words, {\em we find no convincing
evidence that the UV luminosity density decreases with redshift for $z>2$}.

\section{Effects of Cosmological $(1+z)^3$ Surface Brightness Dimming}

 To meaningfully compare the comoving UV luminosity density of the local,
low-redshift Universe with that of the distant, high-redshift Universe, it is
necessary to account for the enormous effect of cosmological $(1+z)^3$
surface brightness (SB) dimming. At $z<0.1$ this effect is less than a factor
of 1.3, but at $z=5$ it becomes a factor of 216. Because low-redshift galaxies
are viewed to
much lower {\em intrinsic} SB thresholds than are high-redshift galaxies,
certain corrections must be applied in order to view all galaxies at a common
intrinsic SB threshold.

  We applied these corrections to our catalog of 1067 galaxies.
First, we
applied monochromatic SB corrections and $K$-corrections on a pixel-by-pixel
basis, as in \cite{B98}, to correct the flux of each galaxy in our catalog to
the value that would be observed if the galaxy were placed at redshift
$z = 5$. We also applied small
corrections to account for the different \wfpc\ passbands that sample
rest-frame 1500 \AA\ at different redshifts.

  Next, we applied a uniform SB cut on a pixel-by-pixel basis, excluding pixels
that failed to meet a minimum SB threshold.  The threshold was determined by
assuming that objects are detected in the F814W image to within $\sim 1 \sigma$
of sky.
Lastly, we reversed the monochromatic SB
corrections and $K$-corrections on a pixel-by-pixel basis to bring each galaxy
back to its original redshift and again calculated the comoving luminosity
density versus redshift.

The results are shown in the bottom panel left of Figure 2, which plots the
comoving UV luminosity density versus redshift of high intrinsic surface
brightness regions.
It is evident that (1) {\em the
comoving UV luminosity density of high intrinsic surface brightness
regions increases by two orders of magnitude from $z\simeq 0$ to $z\simeq 5$},
and (2) star-forming objects
seen at $z > 2.5$ are relatively rare at $z < 2.5$.
Indeed, the righthand panel of Figure 2 also illustrates this point, in
that it shows how the average SB of galaxies in the HDF increases with
redshift.


It is also interesting to note that in the lowest redshift bin
($z=0.0-0.5$) of the lefthand panel of Figure 2 (both top and bottom), where
the SB dimming effect is smallest,
the intrinsically high SB regions make up only 0.5\% of the total UV
luminosity density. If this ratio is the same at high redshifts as it is
at low redshifts then our measurements of the UV luminosity density at high
redshifts in Figure 1{\it a} may need to be increased by up to a factor of
150. Although this ratio may be quite different at high redshifts than it is at
low redshifts, this suggests that the UV luminosity density could be a strongly
increasing function of redshift.

We find that objects with star
formation rates comparable to those at $z$\cge 3 are very rare in the nearby
Universe. This implies that a majority of the star formation may have occurred
at very high redshifts, and therefore that a peak in the
star formation rate density of the Universe has not yet been observed and
may lie somewhere at $z>5$.

\acknowledgements{SMP and KML acknowledge support from NASA grant NAGW-4422 and
NSF grant AST-9624216. AF acknowledges support from a grant from the Australian
Research Council.}

\end{document}